\documentclass[runningheads]{llncs}

\usepackage[utf8]{inputenc}
\usepackage[table]{xcolor}
\usepackage{todonotes}
\usepackage{comment}
\usepackage{url}

\usepackage{xspace}
\usepackage{multirow}
\usepackage{float}
\usepackage{array}
\usepackage{tikz}
\usepackage{threeparttable}
\usepackage{textcomp}

\usepackage{pgfplots}
\pgfplotsset{compat=newest}
\pgfplotsset{plot coordinates/math parser=false}
\usetikzlibrary{plotmarks}
\usetikzlibrary{mindmap,shadows,patterns,calc}
\usepackage[framemethod=tikz]{mdframed}
\definecolor{usethiscolorhere}{rgb}{0.86666,0.78431,0.78431}

\usepackage{amssymb}
\usepackage{booktabs}
\usepackage{pifont}
\usepackage{multirow}
\usepackage{paralist}

\usepackage{graphicx}
\usepackage{caption}
\usepackage{wrapfig}
\usepackage{xcolor}
\usepackage{subcaption}

\usepackage[ruled,norelsize]{algorithm2e}
\makeatletter
\newcommand{\removelatexerror}{\let\@latex@error\@gobble}
\makeatother

\usepackage{listings}
\DeclareCaptionLabelFormat{alglabel}{Alg.~#2}

\usepackage[final,bookmarks=false]{hyperref}
\usepackage{worldflags}

\let\llncssubparagraph\subparagraph
\let\subparagraph\paragraph
\usepackage{titlesec}

\titlespacing*{\section}
{0pt}{2.0ex plus 1ex minus .2ex}{1.0ex plus .2ex minus .2ex}
\titlespacing*{\subsection}
{0pt}{.8ex plus .2ex minus .2ex}{.4ex plus .2ex minus .2ex}
\titlespacing*{\subsubsection}
{0pt}{.4ex plus .2ex minus .2ex}{.2ex plus .2ex minus .2ex}
\titlespacing*{\paragraph}
{0pt}{.6ex plus .2ex minus .2ex}{.3ex plus .2ex minus .2ex}

\let\subparagraph\llncssubparagraph

\usepackage{amsmath}

\usepackage{cleveref}
\Crefname{section}{Sect.}{Sects.}
\Crefname{figure}{Fig.}{Figs.}
\Crefname{table}{Tab.}{Tabs.}

\usepackage{caption}
\usepackage{setspace}

\begin{document}

\title{Enabling mixed-precision with the help of tools:\\ A Nekbone case study}

\author{Yanxiang Chen\inst{1}, Pablo de Oliveira Castro\inst{2}, Paolo Bientinesi\inst{1}, and Roman Iakymchuk\inst{1,3}}
\authorrunning{Y~Chen, P~de Oliveira Castro, P~Bientinesi, R~Iakymchuk}

\institute{
Ume\aa\  University, Sweden.
\email{firstname.lastname@umu.se}
\and
Université Paris-Saclay, UVSQ, LI-PaRAD, France. 
\email{pablo.oliveira@uvsq.fr}
\and
Uppsala University, Sweden.
}

\maketitle

\begin{abstract}
\vspace{-15pt}
Mixed-precision computing has the potential to significantly reduce the cost of exascale computations, but determining when and how to implement it in programs can be challenging.
In this article, we consider Nekbone, a mini-application for the CFD solver Nek5000, as a case study, and  propose a methodology for enabling mixed-precision with the help of computer arithmetic tools and roofline model. We evaluate the derived mixed-precision program by combining metrics in three dimensions: accuracy, time-to-solution, and energy-to-solution. Notably, the introduction of mixed-precision in Nekbone, reducing time-to-solution by $40.7\%$ and energy-to-solution by $47\%$ on 128 MPI ranks.
\vspace{-5pt}
\end{abstract}

\section{Introduction}
With the advent of exascale computing, delivering $10^{18}$ operations per second, there is a great effort to make applications and solvers run faster and efficiently utilize HPC systems. Initially, such applications and solvers were parallelized to tackle bigger problems and speed up computations. Programming models were of good help in this effort by enhancing the underlying communication as well as by making communication asynchronous. As a side effect, parallelization and modification of the classic algorithms, e.g. Krylov-type solvers~\cite{Saa03}, led to partial violation of numerical properties that were re-established by a combined numerical analysis and computer arithmetic effort.
The task is not trivial and requires expertise and skills in different fields. The origin of this problem lies in finite-precision floating-point operations that are commutative but non-associative due to rounding errors. 
For example, denoting $\oplus$ the addition in the double-precision floating-point arithmetic,
$(-1 \oplus 1) \oplus
2^{-53} \neq -1 \oplus (1 \oplus 2^{-53})$ since $(-1 \oplus 1) \oplus
2^{-53} = 2^{-53}$ and $-1 \oplus (1 \oplus
2^{-53} )=0$.
Thus, the accuracy and stability of numerical algorithms rely on error estimates in finite precision~\cite{Hig02}. 

The energy consumption constraint for large-scale computing encourages scientists to revise the architecture design of hardware, linear algebra algorithms, and applications. The main idea is to make the computing cost sustainable and apply the \emph{lagom} principle (in Swedish: just the right amount), especially regarding working and storage precision. The gain in reducing and mixing precision brings not only faster time-to-solution but also a better energy footprint. Applications are relatively slow in picking the trend of energy-efficient computing due to their long-standing development (often over decades) and both complex and sophisticated code with thousands if not millions of lines of code. Many applications share one thing in common: 80\,\% of their execution time is spent in 20\,\% of their code. In this article, we create a bridge between algorithmic development in mixed-precision, often rooted in numerical linear algebra, and applications with the help of tools. Our main contributions are as follows:
\begin{itemize}
    \item As application developers are familiar with profiling and tracing tools used for performance optimisation, we want to introduce them to computer arithmetic tools that can be used for floating-point precision reduction.
    \item With this in mind, we analyze Nekbone with the Verificarlo tool and identify potentials for precision cropping. Additionally, we use Monte Carlo Arithmetic in order to simulate fluctuation in floating-point computations and evaluate the accuracy of reduced precision computations.
    \item We carefully mix double and single precision, where the solver runs exclusively in single, for computations of two common examples, resulting in up to 41\,\% reduction in execution time and 47\,\% in energy-to-solution. 
\end{itemize}

The rest of this article is organized as follows: \Cref{sec:background} describes floating-point arithmetic, Verificarlo, and roofline model. \Cref{sec:methodology} introduces our methodology that builds on three pillars: 1/ code inspection with Verificarlo in~\Cref{sec:inspection} and roofline modeling in~\Cref{sec:roofline}; 2/ strategy to enable mixed-precision in~\Cref{sec:mixed}; 3/ validation and verification in applications~\Cref{sec:results}. Finally, \Cref{sec:conclusion} discusses the main outcomes and outlines future work.

\section{Background}
\label{sec:background}
At first, we will brief on {\bf floating-point (FP) arithmetic} that
consists in approximating real numbers by numbers that have a finite, fixed-precision representation adhering to the IEEE 754 standard. For instance, a FP number is represented on computers with a significand, an exponent, and a sign:
\vspace*{-3mm}
$$x = \pm \underbrace{x_0 . x_1 \ldots x_{M-1}}_{mantissa} \times b^{e}, \,\, 0 \leq x_i \leq b-1, \,\, x_0 \neq 0,$$\\[-4mm]
where $b$ is the  basis ($2$ in our case), $M$ is the precision, and $e$ stands for the exponent, i.e. range. 
The IEEE 754 standard requires correctly rounded results for the basic arithmetic operations $(+, -, \times , /, \sqrt{~},$ {\tt fma}$)$. It 
means that these operations are performed as if the result was first computed with an infinite precision and then rounded to the floating-point format. The correct rounding criterion guarantees a unique, deterministic, and well-defined answer.

{\bf Verificarlo}~\cite{Denis2016verificarlo} is an open-source tool, built upon the LLVM compiler, to analyze and optimise floating point computation in large programs.  Verificarlo is available at \url{http://www.github.com/verificarlo/verificarlo}. Verificarlo replaces at compilation time each floating point operation by a
custom call. After compilation, the program can run with various backends to explore FP issues and optimisations. Verificarlo instruments FP operations at the optimised Intermediate
Representation level (IR). This provides two main advantages: 1) Verificarlo can operate on any source language supported by the LLVM ecosystem including C and C++ through {\tt clang}, and Fortran through {\tt flang}; 2) Verificarlo instrumentation operates after all the other front-end and middle-end optimisation passes capturing most compiler effects on FP operations. 
The two major backends are the Variable Precision (VPREC) backend and the Monte Carlo Arithmetic (MCA) backend.

\textbf{VPREC backend}~\cite{Chatelain2019automatic} transparently emulates lower precisions that fit into the original FP type. For example, if the original program uses {\tt binary64} (double) numbers, the user can emulate FP formats with a pseudo-mantissa of size $t \in [1, 52]$  and an exponent of size $r \in [2, 11]$.  VPREC intercepts each FP operation, performs the computation in double precision and rounds the result to the emulated precision and range. VPREC has been carefully designed to correctly handle overflow, underflows, and denormals.

\textbf{MCA backend} implements different stochastic rounding modes that can be used to estimate the effect and propagation of numerical errors in large programs.  MCA can simulate the effect of different FP precisions by operating at a virtual precision $t$. To model errors on a value $x$, MCA uses the noise function $\text{inexact}(x) = x + 2^{e_x-t}\xi$,
where $e_x = \lfloor \log_2 \left|x\right| \rfloor +1 $ is the order of magnitude of $x$ and $\xi$ is a uniformly distributed random variable in the range $\left(-\frac{1}{2},\frac{1}{2}\right)$. During the MCA run of a given program,
the result and/ or operands of each FP operation is replaced by a perturbed computation modeling the losses of accuracy~\cite{parker97}.
In this study, we use two MCA variants that correspond to different substitutions of a FP operation \mbox{$y\circ z$} where $\circ \in \{+, -, *, / \}$: 1/ Random Rounding (RR) only introduces perturbation on the output \(\text{round}(\text{inexact}(y\circ z))\); 2/ full MCA (MCA) introduces perturbation on operands and the result \(\text{round}(\text{inexact}(\text{inexact}(y)\circ \text{inexact}(z)))\). From a set of MCA samples, it is possible to estimate~\cite{Sohier2021Confidence} the significant bits $s_2$  of a computation, $s_{2} = -\log_{2}\left|{\frac{ \sigma}{\mu}}\right|$, where $\sigma$ and $\mu$ are respectively the standard deviation and mean of the samples.

In addition, we use the {\bf roofline model}~\cite{roofline} that provides a visual representation that integrates floating-point performance, operational or arithmetic intensity, and memory performance in a two-dimensional graph. 

Mixed-precision reduces the computation cost of arithmetic operations and also the communication and memory bandwidth. Depending on the code characteristics, mixed-precision gains are going to come from improving computation, bandwidth, or both. Because the roofline evaluates bottlenecks in both resources, it is a good tool for modeling mixed-precision effects.

\section{Methodology}
\label{sec:methodology}
\begin{figure}[!ht]
\vspace*{-7mm}
\centering
\begin{tikzpicture}[node distance=1.0cm]
  \node (start) {\textbf{Original application}};
  \node (speedup) [below of=start, draw=black] {Check speed-up (Roofline model)};
  \node (accuracy) [below of=speedup, draw=black] {Check accuracy (Verificarlo)};
  \node (candidate) [below of=accuracy, xshift=3cm] {\textbf{Candidate code-sections}};
  \node (port) [above of=candidate, xshift=3cm] {\fbox{Port algorithms to mixed-precision}};
  \node (validation) [above of=port, draw=black] {Expert Validation};
  \node (mixed) [above of=validation] {\textbf{Mixed-Precision application}};

  \draw [->] (start) -> (speedup);
  \draw [->] (speedup) -> (accuracy);
  \draw [->] (accuracy) -> (candidate);
  \draw [->] (candidate) -> (port);
  \draw [->] (port) -> (validation);
  \draw [->] (validation) -> (mixed);
\end{tikzpicture}
\vspace*{-2mm}
\caption{Methodology flow-chart. \label{fig:flow}}
\vspace*{-7mm}
\end{figure}
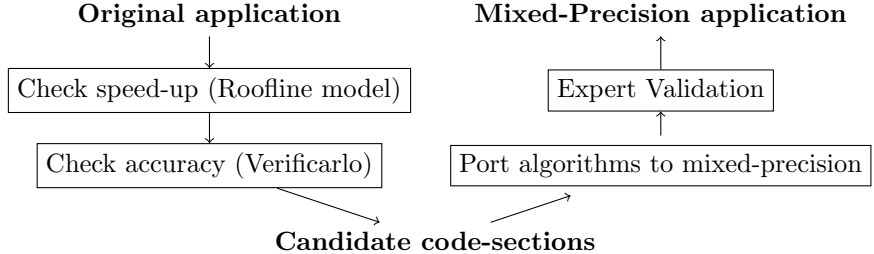

Adapting an application to use mixed-precision is a costly process that is currently hard to automate. It requires manual modification of data-structures types and both mathematical and communication library calls.
We propose a methodology, captured in~\Cref{fig:flow}, to assess which parts of a program can benefit from such work with the help of tools. Candidate code sections must:\\[-6mm]
\begin{itemize}
    \item achieve a significant speed-up when using lower, eg {\tt binary32} (single), data types and operations (speed-up check);
    \item achieve required accuracy with lower-precision operations (accuracy check).    
\end{itemize}
\vspace*{-2mm}

In the first step, we profile the code sections and produce roofline models, which we use to evaluate the potential speed-up of moving computation and communications to lower precision.
In the second step, we check the accuracy with Verificarlo VPREC and MCA backends. The backends estimate the error introduced by lower, eg {\tt binary32}, precision on the remaining code sections.  
VPREC is used first on each section separately. The forward error of the output is monitored. If it raises above a user's defined threshold, the section is filtered out.
The set of candidate sections is further pruned thanks to the MCA backend, which eliminates sections that are unstable with stochastic rounding. 

Regarding accuracy, this methodology does not give formal guarantees on the error for arbitrary datasets. However, by using different datasets in the pruning step, the user can efficiently eliminate code sections that are too sensitive to single-precision computations. Similarly, the roofline model gives a quick and efficient method to prune code-sections with low return-on-investment.
Our contribution consists, therefore, in providing an automatic and affordable methodology for reducing the scope of code considered for porting to lower precision.

After the automatic pruning, the expert should modify the code to use lower precision when needed and validate the new algorithm error guarantees. This step follows with time-to-solution and energy-to-solution measurements.
In the following sections, we demonstrate this methodology on the Nekbone mini-app.

\section{Inspecting precision appetites with VerifiCarlo}
\label{sec:inspection}

\subsection{Nekbone}
Nekbone~\cite{nekbone} is a mini-app capturing the basic structure and design of the extensive Nek5000~\cite{nek5000} software, which is a high-order, incompressible Navier-Stokes solver based on the spectral element method. Nekbone solves a standard Poisson equation using the Conjugate Gradient (CG) method with a simple multigrid preconditioner (the preconditioner is optional when compiling). The computational domain is partitioned into high-order quadrilateral elements.

The solution phase performs multiple conjugate gradient iterations invoking the main computational kernel. Overall, each iteration consists of vector operations, matrix-matrix multiply operations, nearest-neighbor communications with MPI, and MPI Allreduce operations. The main computational kernel (CG loop) in Nekbone is written in Fortran 77, and the Gather-Scatter library~\cite{gslib} (an individual component of Nek5000) for nearest-neighbour communications is written in C. 
Although both double and single precision data types are technically compatible with some of Nekbone's components, Nekbone is developed entirely in double precision, and neither single nor mixed precision is supported.

\Cref{tab:cg-profiling} shows the code for the CG loop, the operations associated with each line, and the profiling results, which consist of {\em include time} and {\em self time}. Include time is the percentage of the overall runtime, while self time is the percentage of the runtime. If a kernel has no callees, then the include time is equal to the self time. If a kernel does not perform any calculation then the self time is $0.00\%$.

\begin{figure}
    \vspace*{-13mm}
    \centering
    \resizebox{0.8\linewidth}{!}{\tikzset{every picture/.style={line width=0.75pt}} 

\begin{tikzpicture}[x=0.75pt,y=0.75pt,yscale=-1,xscale=1]

\draw    (75.25,85.5) -- (128.25,85.5) -- (128.25,109.5) -- (75.25,109.5) -- cycle  ;
\draw (101.75,97.5) node  [font=\large] [align=left] {\begin{minipage}[lt]{33.32pt}\setlength\topsep{0pt}
\begin{center}
MAIN
\end{center}

\end{minipage}};
\draw    (154.56,85.5) -- (181.56,85.5) -- (181.56,109.5) -- (154.56,109.5) -- cycle  ;
\draw (168.06,97.5) node  [font=\large] [align=left] {\begin{minipage}[lt]{15.65pt}\setlength\topsep{0pt}
\begin{center}
cg
\end{center}

\end{minipage}};
\draw    (211.79,55.5) -- (261.79,55.5) -- (261.79,79.5) -- (211.79,79.5) -- cycle  ;
\draw (236.79,67.5) node  [font=\large] [align=left] {\begin{minipage}[lt]{31.3pt}\setlength\topsep{0pt}
\begin{center}
glsc3
\end{center}

\end{minipage}};
\draw    (213.7,113.25) -- (240.7,113.25) -- (240.7,137.25) -- (213.7,137.25) -- cycle  ;
\draw (227.2,125.25) node  [font=\large] [align=left] {\begin{minipage}[lt]{15.65pt}\setlength\topsep{0pt}
\begin{center}
ax
\end{center}

\end{minipage}};
\draw    (277.33,113.25) -- (324.33,113.25) -- (324.33,137.25) -- (277.33,137.25) -- cycle  ;
\draw (300.83,125.25) node  [font=\large] [align=left] {\begin{minipage}[lt]{29.27pt}\setlength\topsep{0pt}
\begin{center}
ax\_e
\end{center}

\end{minipage}};
\draw    (295.98,55.5) -- (396.98,55.5) -- (396.98,79.5) -- (295.98,79.5) -- cycle  ;
\draw (346.48,67.5) node  [font=\large] [align=left] {\begin{minipage}[lt]{66.01pt}\setlength\topsep{0pt}
\begin{center}
local\_grad3
\end{center}

\end{minipage}};
\draw    (353.88,112.38) -- (469.88,112.38) -- (469.88,136.38) -- (353.88,136.38) -- cycle  ;
\draw (411.88,124.38) node  [font=\large] [align=left] {\begin{minipage}[lt]{76.22pt}\setlength\topsep{0pt}
\begin{center}
local\_grad3\_t
\end{center}

\end{minipage}};
\draw    (505.75,55.5) -- (552.75,55.5) -- (552.75,79.5) -- (505.75,79.5) -- cycle  ;
\draw (529.25,67.5) node  [font=\large] [align=left] {\begin{minipage}[lt]{29.23pt}\setlength\topsep{0pt}
\begin{center}
mxm
\end{center}

\end{minipage}};
\draw    (504.75,112.38) -- (552.75,112.38) -- (552.75,136.38) -- (504.75,136.38) -- cycle  ;
\draw (528.75,124.38) node  [font=\large] [align=left] {\begin{minipage}[lt]{29.95pt}\setlength\topsep{0pt}
\begin{center}
add2
\end{center}

\end{minipage}};
\draw    (128.25,97.5) -- (152.56,97.5) ;
\draw [shift={(154.56,97.5)}, rotate = 180] [fill={rgb, 255:red, 0; green, 0; blue, 0 }  ][line width=0.08]  [draw opacity=0] (12,-3) -- (0,0) -- (12,3) -- cycle    ;
\draw    (181.56,103.83) -- (211.89,118.07) ;
\draw [shift={(213.7,118.92)}, rotate = 205.14] [fill={rgb, 255:red, 0; green, 0; blue, 0 }  ][line width=0.08]  [draw opacity=0] (12,-3) -- (0,0) -- (12,3) -- cycle    ;
\draw    (181.56,91.61) -- (209.96,79.21) ;
\draw [shift={(211.79,78.41)}, rotate = 156.42] [fill={rgb, 255:red, 0; green, 0; blue, 0 }  ][line width=0.08]  [draw opacity=0] (12,-3) -- (0,0) -- (12,3) -- cycle    ;
\draw    (240.7,125.25) -- (275.33,125.25) ;
\draw [shift={(277.33,125.25)}, rotate = 180] [fill={rgb, 255:red, 0; green, 0; blue, 0 }  ][line width=0.08]  [draw opacity=0] (12,-3) -- (0,0) -- (12,3) -- cycle    ;
\draw    (310.31,113.25) -- (335.75,81.07) ;
\draw [shift={(336.99,79.5)}, rotate = 128.33] [fill={rgb, 255:red, 0; green, 0; blue, 0 }  ][line width=0.08]  [draw opacity=0] (12,-3) -- (0,0) -- (12,3) -- cycle    ;
\draw    (324.33,125.06) -- (351.88,124.85) ;
\draw [shift={(353.88,124.83)}, rotate = 179.55] [fill={rgb, 255:red, 0; green, 0; blue, 0 }  ][line width=0.08]  [draw opacity=0] (12,-3) -- (0,0) -- (12,3) -- cycle    ;
\draw    (396.98,67.5) -- (503.75,67.5) ;
\draw [shift={(505.75,67.5)}, rotate = 180] [fill={rgb, 255:red, 0; green, 0; blue, 0 }  ][line width=0.08]  [draw opacity=0] (12,-3) -- (0,0) -- (12,3) -- cycle    ;
\draw    (436.64,112.38) -- (503.95,79.76) ;
\draw [shift={(505.75,78.89)}, rotate = 154.15] [fill={rgb, 255:red, 0; green, 0; blue, 0 }  ][line width=0.08]  [draw opacity=0] (12,-3) -- (0,0) -- (12,3) -- cycle    ;
\draw    (469.88,124.38) -- (502.75,124.38) ;
\draw [shift={(504.75,124.38)}, rotate = 180] [fill={rgb, 255:red, 0; green, 0; blue, 0 }  ][line width=0.08]  [draw opacity=0] (12,-3) -- (0,0) -- (12,3) -- cycle    ;

\end{tikzpicture}}
    \caption{Call graph of Nekbone without preconditioner, only the most time-consuming computational kernels are shown.}
    \label{fig:callgraph}
    \vspace*{-7mm}
\end{figure}
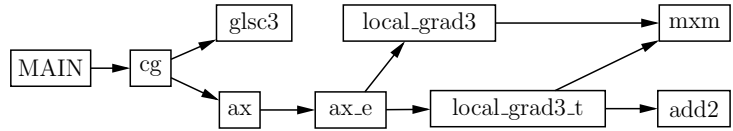

\begin{figure}
    \vspace*{-2mm}
    \centering
    \resizebox{\linewidth}{!}{\tikzset{every picture/.style={line width=0.75pt}} 

\begin{tikzpicture}[x=0.75pt,y=0.75pt,yscale=-1,xscale=1]

\draw    (27.25,98.5) -- (80.25,98.5) -- (80.25,122.5) -- (27.25,122.5) -- cycle  ;
\draw (53.75,110.5) node  [font=\large] [align=left] {MAIN};
\draw    (112.42,98.5) -- (139.42,98.5) -- (139.42,122.5) -- (112.42,122.5) -- cycle  ;
\draw (125.92,110.5) node  [font=\large] [align=left] {cg};
\draw    (167.26,68.5) -- (232.26,68.5) -- (232.26,92.5) -- (167.26,92.5) -- cycle  ;
\draw (199.76,80.5) node  [font=\large] [align=left] {solvem};
\draw    (168.42,125.25) -- (195.42,125.25) -- (195.42,149.25) -- (168.42,149.25) -- cycle  ;
\draw (181.92,137.25) node  [font=\large] [align=left] {ax};
\draw    (199.77,183.25) -- (246.77,183.25) -- (246.77,207.25) -- (199.77,207.25) -- cycle  ;
\draw (223.27,195.25) node  [font=\large] [align=left] {ax\_e};
\draw    (283.28,125.25) -- (384.28,125.25) -- (384.28,149.25) -- (283.28,149.25) -- cycle  ;
\draw (333.78,137.25) node  [font=\large] [align=left] {local\_grad3};
\draw    (270.04,183.38) -- (386.04,183.38) -- (386.04,207.38) -- (270.04,207.38) -- cycle  ;
\draw (328.04,195.38) node  [font=\large] [align=left] {local\_grad3\_t};
\draw    (440.75,146.5) -- (487.75,146.5) -- (487.75,170.5) -- (440.75,170.5) -- cycle  ;
\draw (464.25,158.5) node  [font=\large] [align=left] {mxm};
\draw    (257.94,69.5) -- (386.94,69.5) -- (386.94,93.5) -- (257.94,93.5) -- cycle  ;
\draw (322.44,81.5) node  [font=\large] [align=left] {h1mg\_schwarz};
\draw    (414.45,92) -- (507.45,92) -- (507.45,116) -- (414.45,116) -- cycle  ;
\draw (460.95,104) node  [font=\large] [align=left] {h1mg\_fdm};
\draw    (414.45,42.5) -- (537.45,42.5) -- (537.45,66.5) -- (414.45,66.5) -- cycle  ;
\draw (475.95,54.5) node  [font=\large] [align=left] {h1mg\_extrude};
\draw    (529.04,92) -- (651.04,92) -- (651.04,116) -- (529.04,116) -- cycle  ;
\draw (590.04,104) node  [font=\large] [align=left] {h1mg\_do\_fast};
\draw    (521.29,146.5) -- (658.29,146.5) -- (658.29,170.5) -- (521.29,170.5) -- cycle  ;
\draw (589.79,158.5) node  [font=\large] [align=left] {h1mg\_tnsr3d\_el};
\draw    (439.75,183.38) -- (487.75,183.38) -- (487.75,207.38) -- (439.75,207.38) -- cycle  ;
\draw (463.75,195.38) node  [font=\large] [align=left] {add2};
\draw    (80.25,110.5) -- (110.42,110.5) ;
\draw [shift={(112.42,110.5)}, rotate = 180] [fill={rgb, 255:red, 0; green, 0; blue, 0 }  ][line width=0.08]  [draw opacity=0] (12,-3) -- (0,0) -- (12,3) -- cycle    ;
\draw    (139.42,116.95) -- (166.62,129.94) ;
\draw [shift={(168.42,130.8)}, rotate = 205.53] [fill={rgb, 255:red, 0; green, 0; blue, 0 }  ][line width=0.08]  [draw opacity=0] (12,-3) -- (0,0) -- (12,3) -- cycle    ;
\draw    (139.42,105.01) -- (168.37,93.25) ;
\draw [shift={(170.22,92.5)}, rotate = 157.89] [fill={rgb, 255:red, 0; green, 0; blue, 0 }  ][line width=0.08]  [draw opacity=0] (12,-3) -- (0,0) -- (12,3) -- cycle    ;
\draw    (190.48,149.25) -- (213.55,181.62) ;
\draw [shift={(214.71,183.25)}, rotate = 234.52] [fill={rgb, 255:red, 0; green, 0; blue, 0 }  ][line width=0.08]  [draw opacity=0] (12,-3) -- (0,0) -- (12,3) -- cycle    ;
\draw    (246.13,183.25) -- (309.15,150.18) ;
\draw [shift={(310.92,149.25)}, rotate = 152.31] [fill={rgb, 255:red, 0; green, 0; blue, 0 }  ][line width=0.08]  [draw opacity=0] (12,-3) -- (0,0) -- (12,3) -- cycle    ;
\draw    (246.77,195.28) -- (268.04,195.3) ;
\draw [shift={(270.04,195.31)}, rotate = 180.07] [fill={rgb, 255:red, 0; green, 0; blue, 0 }  ][line width=0.08]  [draw opacity=0] (12,-3) -- (0,0) -- (12,3) -- cycle    ;
\draw    (384.28,145.48) -- (438.77,154.35) ;
\draw [shift={(440.75,154.67)}, rotate = 189.25] [fill={rgb, 255:red, 0; green, 0; blue, 0 }  ][line width=0.08]  [draw opacity=0] (12,-3) -- (0,0) -- (12,3) -- cycle    ;
\draw    (372.36,183.38) -- (438.82,165.38) ;
\draw [shift={(440.75,164.86)}, rotate = 164.85] [fill={rgb, 255:red, 0; green, 0; blue, 0 }  ][line width=0.08]  [draw opacity=0] (12,-3) -- (0,0) -- (12,3) -- cycle    ;
\draw    (232.26,80.76) -- (255.94,80.96) ;
\draw [shift={(257.94,80.97)}, rotate = 180.47] [fill={rgb, 255:red, 0; green, 0; blue, 0 }  ][line width=0.08]  [draw opacity=0] (12,-3) -- (0,0) -- (12,3) -- cycle    ;
\draw    (386.94,70.16) -- (412.48,65.66) ;
\draw [shift={(414.45,65.32)}, rotate = 170.02] [fill={rgb, 255:red, 0; green, 0; blue, 0 }  ][line width=0.08]  [draw opacity=0] (12,-3) -- (0,0) -- (12,3) -- cycle    ;
\draw    (386.94,91.98) -- (412.48,96.13) ;
\draw [shift={(414.45,96.45)}, rotate = 189.23] [fill={rgb, 255:red, 0; green, 0; blue, 0 }  ][line width=0.08]  [draw opacity=0] (12,-3) -- (0,0) -- (12,3) -- cycle    ;
\draw    (507.45,104) -- (527.04,104) ;
\draw [shift={(529.04,104)}, rotate = 180] [fill={rgb, 255:red, 0; green, 0; blue, 0 }  ][line width=0.08]  [draw opacity=0] (12,-3) -- (0,0) -- (12,3) -- cycle    ;
\draw    (589.98,116) -- (589.85,144.5) ;
\draw [shift={(589.84,146.5)}, rotate = 270.26] [fill={rgb, 255:red, 0; green, 0; blue, 0 }  ][line width=0.08]  [draw opacity=0] (12,-3) -- (0,0) -- (12,3) -- cycle    ;
\draw    (521.29,158.5) -- (489.75,158.5) ;
\draw [shift={(487.75,158.5)}, rotate = 360] [fill={rgb, 255:red, 0; green, 0; blue, 0 }  ][line width=0.08]  [draw opacity=0] (12,-3) -- (0,0) -- (12,3) -- cycle    ;
\draw    (386.04,195.38) -- (437.75,195.38) ;
\draw [shift={(439.75,195.38)}, rotate = 180] [fill={rgb, 255:red, 0; green, 0; blue, 0 }  ][line width=0.08]  [draw opacity=0] (12,-3) -- (0,0) -- (12,3) -- cycle    ;

\end{tikzpicture}}
    \caption{Call graph of Nekbone with the preconditioner enabled.}
    \label{fig:callgraph-precond}
    \vspace*{-6mm}
\end{figure}
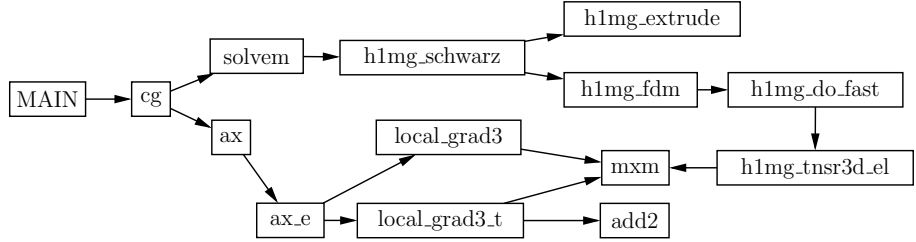

The profiling results are obtained by the Callgrind~\cite{callgrind} tool (a module of Valgrind~\cite{valgrind}); the column no-MGRID shows the timings without multigrid preconditioner. The most time-consuming kernel is $ax$, which calls the $local\_grad3$, $local\_grad3\_t$, $add2$ and $mxm$ (the call graph is shown in~\Cref{fig:callgraph}). Conversely, when the preconditioner is enabled, the kernel $solveM$ is the most time-consuming, followed closely by $ax$
(the call graph is shown in~\Cref{fig:callgraph-precond}).
\begin{table}[!ht]
\vspace*{-10mm}
\caption{Profiling of the CG kernel within Nekbone.} 
\label{tab:cg-profiling}
\begin{threeparttable}[t]
\centering
\begin{tabular}{p{0.41\textwidth}||p{0.177\textwidth}p{0.141\textwidth}p{0.132\textwidth}}
\hline \noalign{\hrule height 0.3pt}
 CG loop\tnote{a} & Description & \multicolumn{2}{l}{Include time, Self time (\%)} \\
\hline
1 \ \textbf{do} iter = 1, miter &  & no MGRID & MGRID \\
\cline{3-4} 
2 \ \ \ \ call \textbf{solveM}(z,r,n) & preconditioner & - & 69.82, 0.00 \\
3 \ \ \ \ rtz1 = \textbf{glsc3}(r,c,z,n) & dot product & 5.79, 5.79 & 1.72, 1.72 \\
4 \ \ \ \ call \textbf{add2s1}(p,z,beta,n) & vector add & 1.76, 1.76 & 0.52, 0.52 \\
5 \ \ \ \ call \textbf{ax}(w,p,g,ur,us,ut,wk,n) & multiplication\tnote{b} & 86.73, 0.00 & 25.83, 0.00 \\
6 \ \ \ \ pap = \textbf{glsc3}(w,c,p,n) & dot product & 5.79, 5.79\tnote{c} & 1.72, 1.72 \\
7 \ \ \ \ call \textbf{add2s2}(x,p,alpha,n) & vector add & 5.05, 5.05 & 1.50, 1.50 \\
8 \ \ \ \ call \textbf{add2s2}(r,w,alphm,n) & vector add & 5.05, 5.05 & 1.50, 1.50 \\
9 \ \ \ \ rtr = \textbf{glsc3}(r,c,r,n) & dot product & 5.79, 5.79 & 1.72, 1.72 \\
10 \ \textbf{enddo} &  & - & - \\
\hline \noalign{\hrule height 0.3pt}
\end{tabular}
\begin{tablenotes}
    \item[a] We include only those lines that contain compute kernels. 
    \item[b] Matrix-matrix multiplications, and accumulations before and after.
    \item[c] Multiple calls to the same kernel are combined, e.g., lines 3\&6.
\end{tablenotes}
\end{threeparttable}
\end{table}

Inspection is a crucial component of our methodology, which is responsible for analyzing precision needs and identifying possibilities for precision cropping as follows\\[-4mm]
\begin{enumerate}
    \item Simulating the application using the VPREC backend with lower precisions and assessing the potential for precision reduction by tracking a few significant variables such as residual, $\alpha$, $\beta$, and $pap$ in the (preconditioned) CG algorithm.
    \item Further verifying convergence sensitivity using stochastic arithmetic through the MCA backend.
\end{enumerate}
\vspace*{-2mm}

We perform the analysis with Verificarlo version 1.0, Flang and LLVM version 14.0.1 (Classic Flang Project~\cite{flang}) installed on Kebnekaise at HPC2N in Sweden. We present more details of Kebnekaise in the experimental part, see~\Cref{sec:results}.

\subsection{Lower precision emulations with VPREC backend}
\begin{figure}[ht]
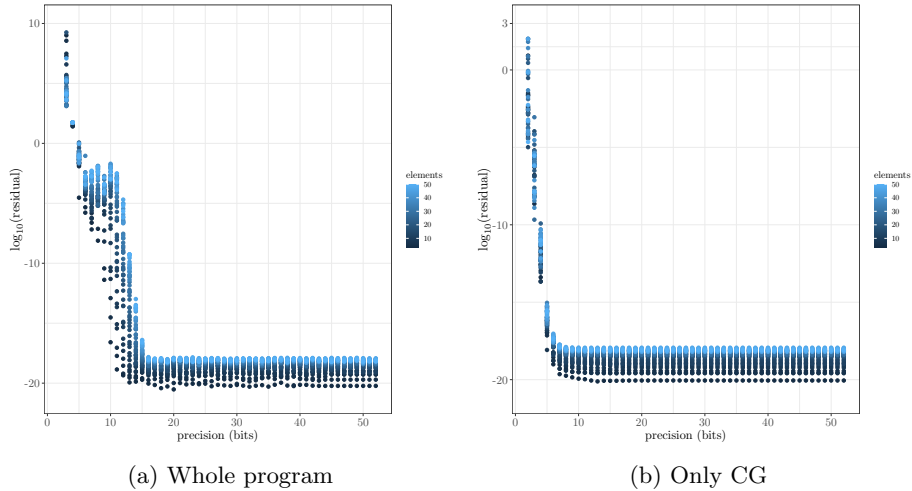

    \vspace*{-6mm} 
    \begin{subfigure}{0.49\textwidth}
        \resizebox{\linewidth}{!}{\input{imgs/rnorm_vprec}}
        \caption{Whole program} \label{fig:vprec-whole}
    \end{subfigure}
    \hspace*{\fill}
    \begin{subfigure}{0.49\textwidth}
        \resizebox{\linewidth}{!}{\input{imgs/rnorm_vprec_cg}}
        \caption{Only CG} \label{fig:vprec-cg}
    \end{subfigure}
    \vspace*{-4mm}
    \caption{Residual with precision (mantissa) from 3 to 52 bits with the VPREC backend for the case without preconditioner.}
    \label{fig:vprec-whole-cg}
    \vspace*{-7mm}
\end{figure}

We use the VPREC backend to instrument double precision FP operations, simulating lower precisions with $t\in [3,52]$ for the whole Nekbone application. \Cref{fig:vprec-whole} illustrates how $residual$ becomes smaller as precision grows, stabilising at $t \geq 16$ at its lowest level. We track the residual because it captures the overall effect of lowering precision on the solution accuracy.

When only the functions involved in the CG kernel are instrumented, the plot becomes even smoother and reaches a stable state at $t \geq 8$, as depicted in~\Cref{fig:vprec-cg}. The results of the VPREC experiments indicate that there is potential for single precision in the entire program. After this initial evaluation, we test the sensitivity of these results with the MCA backend.

\subsection{Sensitivity analysis with the MCA backend}
\begin{figure}[ht]
    \vspace*{-9mm} 
    \begin{subfigure}{0.49\textwidth}
        \includegraphics[width=1.05\linewidth]{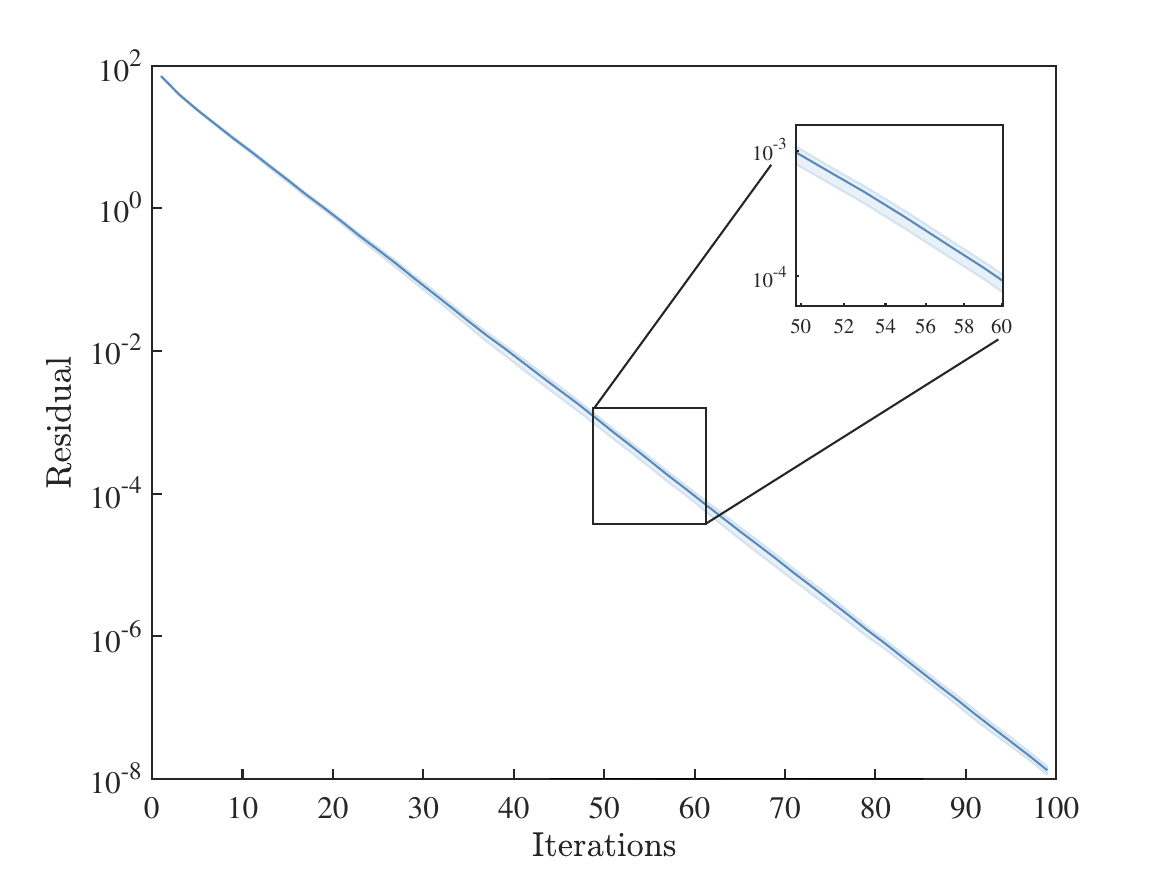}
        \caption{$rr$ mode} \label{fig:mca-rr}
    \end{subfigure}
    \hspace*{\fill}
    \begin{subfigure}{0.49\textwidth}
        \includegraphics[width=1.05\linewidth,height=4.7cm]{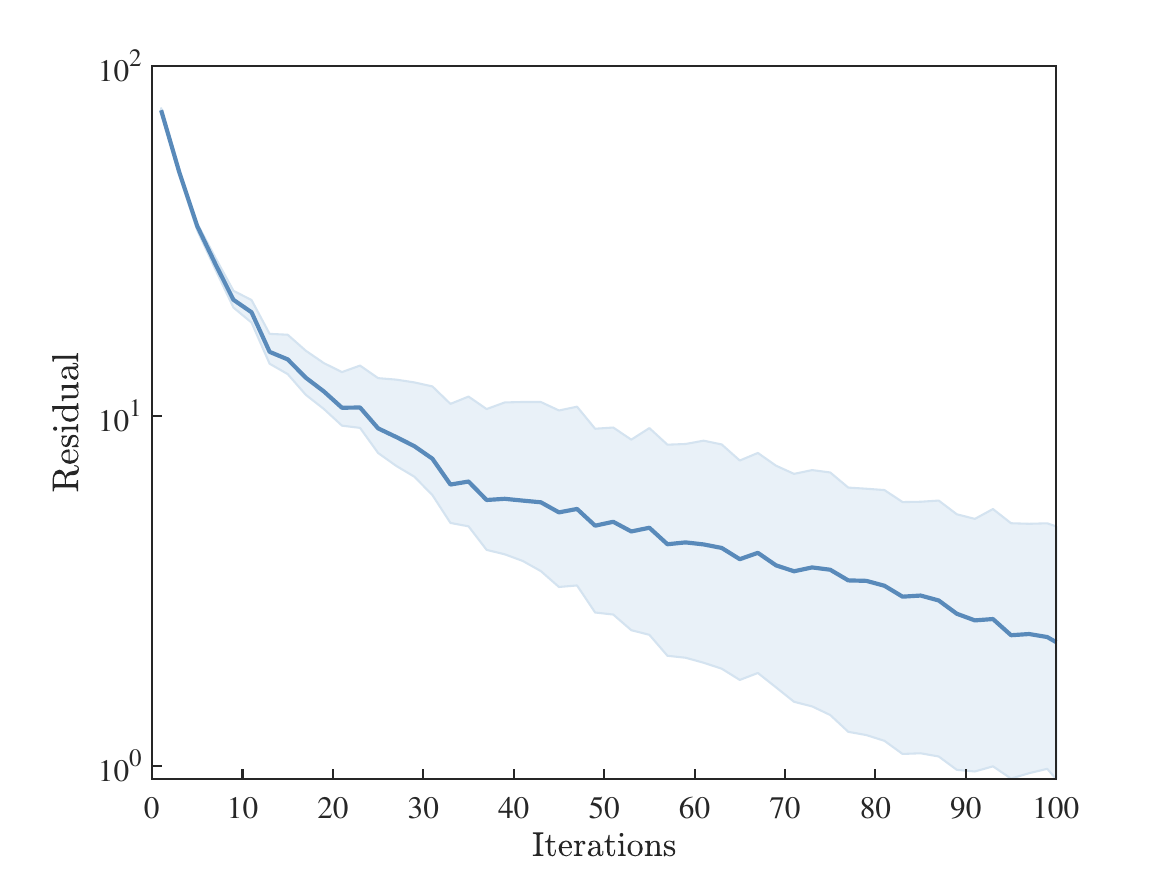}
        \caption{$mca$ mode} \label{fig:mca-mca}
    \end{subfigure}
    \vspace*{-6mm}    
    \caption{Residual history with the MCA backend in the {\em entire Nekbone} mini-app.}
    \label{fig:mca-whole}
    \vspace*{-4mm}    
\end{figure}

We first run the whole program using the MCA backend. The $mca$ and $rr$ modes are employed with the precision $t = 23$ bits. 
We evaluate $20$ MCA runs, to have a statistically significant sample size. \Cref{fig:mca-whole} illustrates the error bar plot (filled area) for $20$ runs, the blue curve represents the mean value $\mu =\sum residual/ 20$ for each iteration, while the upper and lower bounds of the error bar are determined by the maximum and minimum values: $err=[ min,\ max]$.
\begin{figure}[ht]
    \vspace*{-12mm}  
    \begin{subfigure}{0.49\textwidth}
        \includegraphics[width=1.05\linewidth]{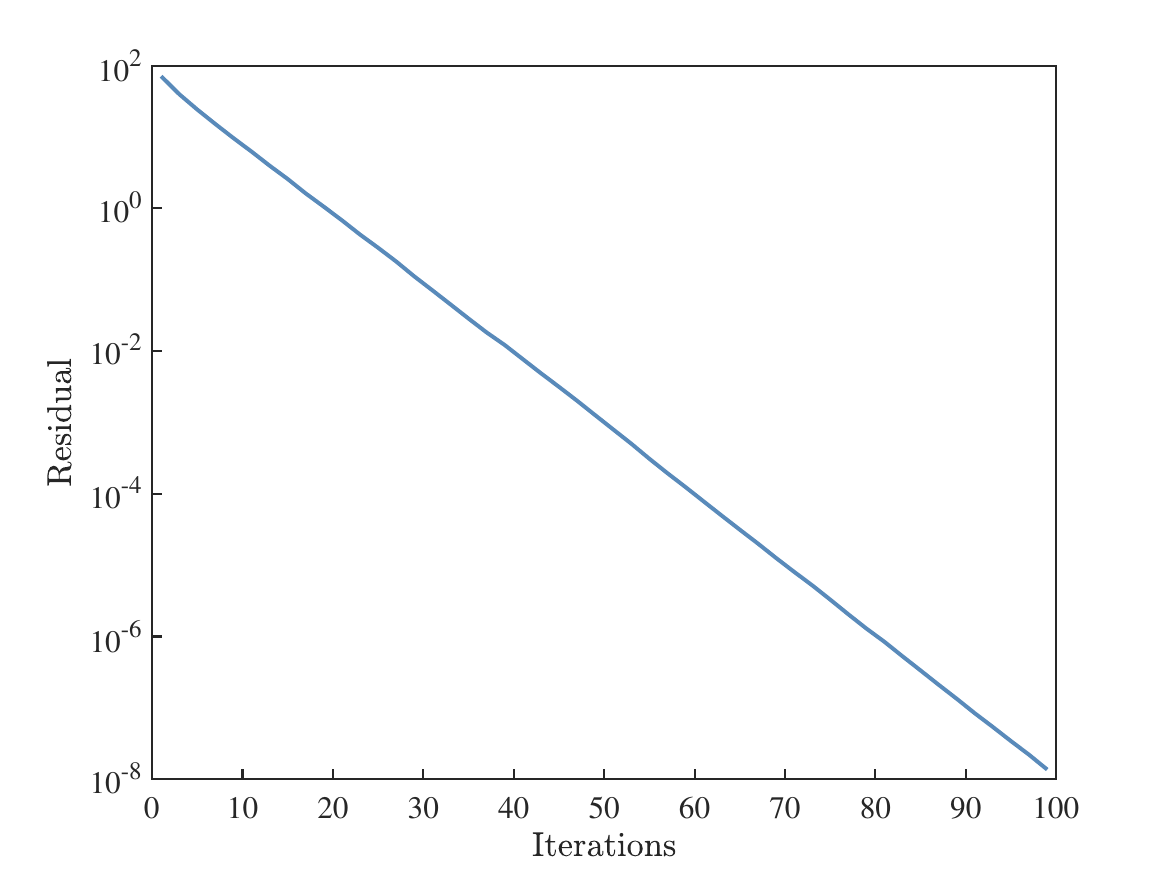}
        \caption{$rr$ mode} \label{fig:mca-rr-onlycg}
    \end{subfigure}
    \hspace*{\fill}
    \begin{subfigure}{0.49\textwidth}
        \includegraphics[width=1.05\linewidth]{imgs/mca_mca_onlycg.pdf}
        \caption{$mca$ mode} \label{fig:mca-mca-onlycg}
    \end{subfigure}
    \vspace*{-6mm}   
    \caption{
    Residual history with the MCA backend in the {\em CG loop only}.}  
    \label{fig:mca-onlycg}
    \vspace*{-7mm}    
\end{figure}

We observe the error surrounding the mean value is negligible in the Random Rounding ($rr$) mode. Hence, the simulation in single precision yields satisfactory results. However, there are large fluctuations with $mca$ mode -- the error accumulates over the course of computation and stops convergence. Subsequently, we conduct a thorough analysis of each kernel (or subroutine/ function) separately using the $mca$ mode in order to identify sources of the fluctuations. 
One particularly sensitive function is {\tt pnormj},  which is called during the initialization phase of Nekbone. Despite the exclusion of this function, the fluctuations remain with the $mca$ mode. 
Yet, if we exclude all initialization functions from the $mca$ instrumentation, the fluctuations stop. The problem seems, therefore, to point to sensitive FP operations occuring at the initialization.

Taking into account this information as well as the profiling results that identify the CG loop as the most time-consuming part of Nekbone, we decide to exclusively concentrate on the CG loop for exploring precision cropping. \Cref{fig:mca-onlycg} depicts the results of our inspection with the $mca$ and $rr$ modes in the CG loop. These results show no fluctuations. 

\begin{wrapfigure}{r}{0.5\textwidth}
    \vspace*{-8mm}
    \centering
    \includegraphics[width=\linewidth]{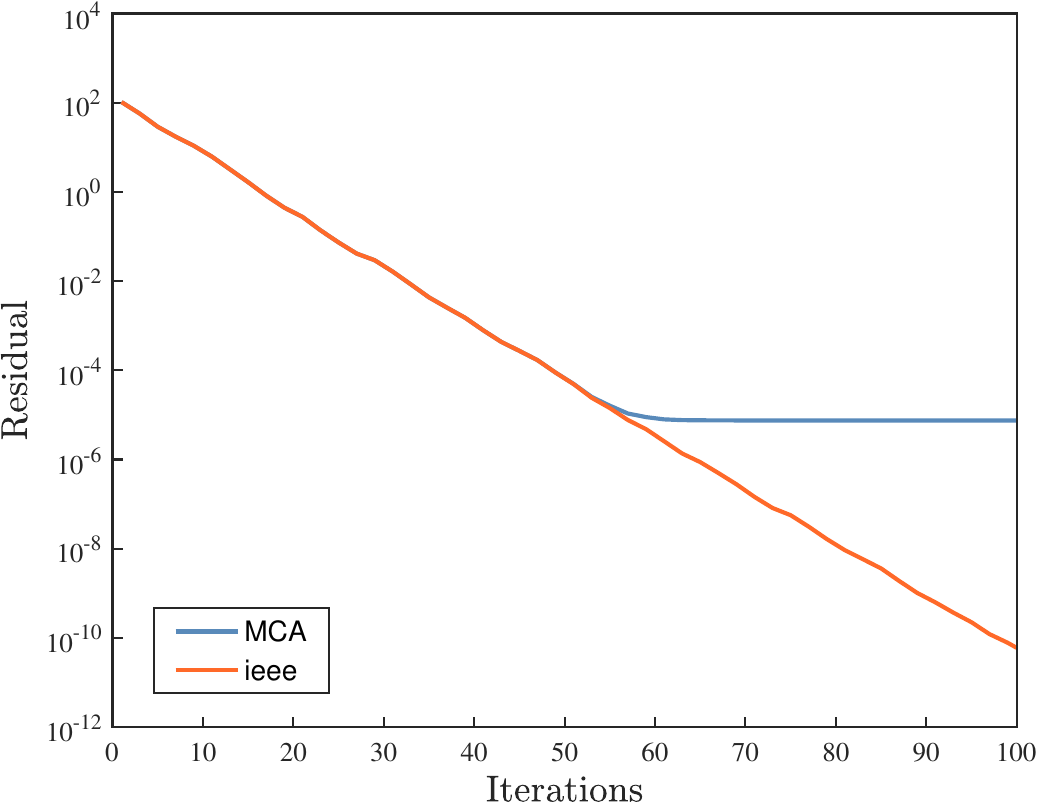}
    \caption{Nekbone with the multigrid preconditioner enabled using both the $rr$ and $mca$ modes in the MCA backend; $ieee$ stands for double precision results.}
    \label{fig:mca-precond}
    \vspace*{-8mm}    
\end{wrapfigure}
Thus, the code inspection with both VPREC and MCA backends reveals a possibility for precision cropping from double to single in the CG loop. In particular, the MCA backend with both modes, introducing noise as an input and output in FP operations, tests the numerical sensitivity of such cropping and confirms this possibility. Certainly, different boundary/initial conditions, different domains, different mashing, or larger data sets may impact the choice of precision. Hence, it is necessary to conduct additional testing, e.g., when the simulation changes, in conjunction with an actual implementation to validate these findings.

In addition, we inspect Nekbone with the multigrid preconditioner enabled. We notice stagnation in both the $rr$ and $mca$ modes using MCA backend, the residuals flattens after the $61$st iteration (with the residual of $7.94\times 10^{-6}$), as depicted in~\Cref{fig:mca-precond}. These results demonstrate that static precision cropping is not applicable with the preconditioner enabled. Further investigation is needed to understand this limitation.

\section{Roofline modeling}
\label{sec:roofline}
The roofline model provides us with valuable insights for assessing and enhancing software for FP computations. This paper aims to identify the bottlenecks of the double precision Nekbone and evaluate the improvements achieved through mixed-precision using the roofline model for analysis. The model is generated with the help of Intel\textsuperscript{\textregistered} Advisor~\cite{advisor} version 2023.2.0 installed on Kebnekaise. We use Intel Compiler Toolkits version 2021.9.0, specifically {\tt ifort} and {\tt icc} compiler, with the {\tt -O2} optimization flag for the code compilation. Note that many optimizations including vectorisation can be enabled with {\tt -O2}.

\begin{figure}[!ht]
    \vspace*{-8mm}
    \centering
    \includegraphics[width=0.95\textwidth]{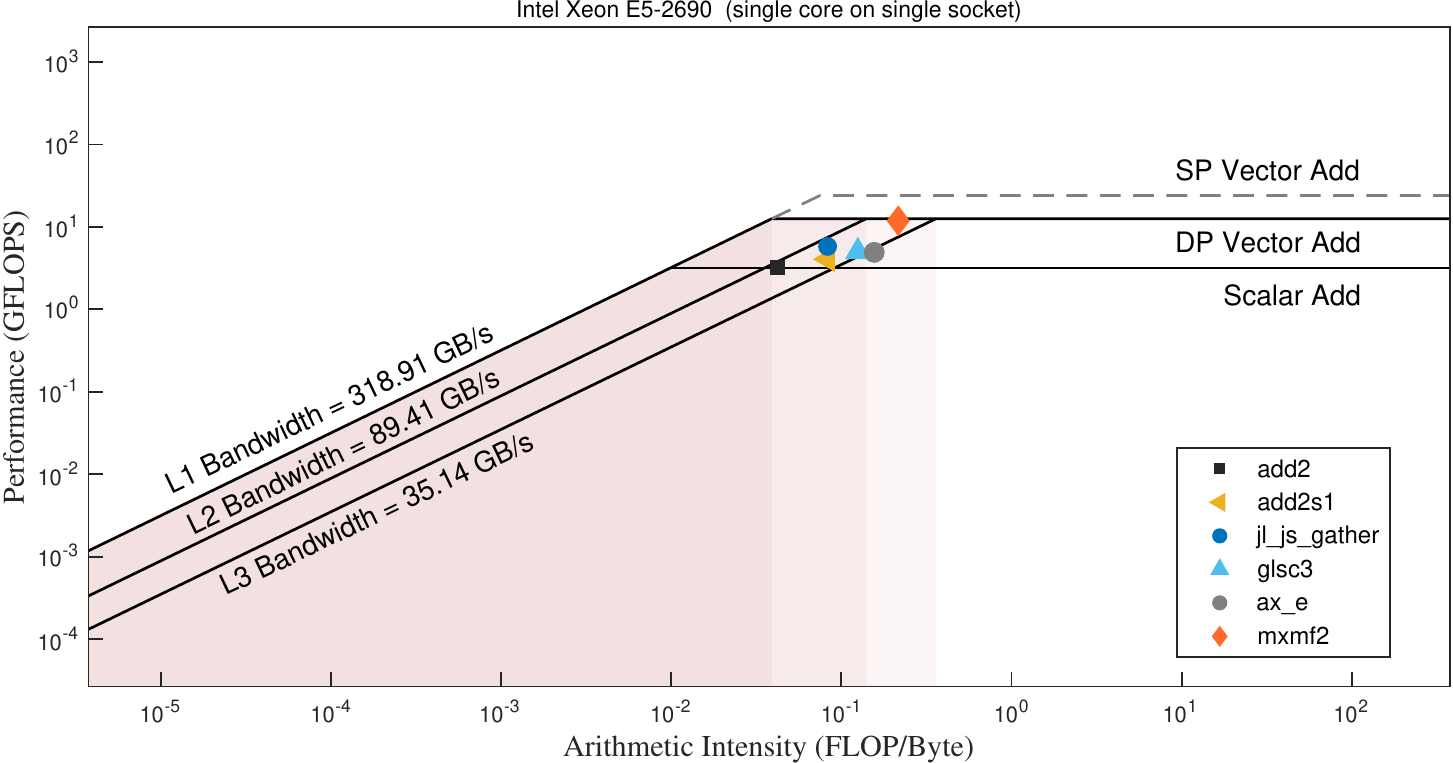}
    \caption{Roofline model for double precision Nekbone.}
    \label{fig:roofline-double}
    \vspace*{-8mm}    
\end{figure}

\Cref{fig:roofline-double} shows the roofline model for the case without preconditioner in double precision on an Intel Xeon E5-2690 processor (single core), with the elements from 1 to 100. The graph is plotted on a logarithmic scale, with three parallel diagonal lines representing the memory bounds Level (L1), L2, and L3 caches. The x-axis represents the arithmetic/ operational intensity (FLOP/ Byte), while the y-axis is the attainable floating-point performance, measured in GFLOPS. Horizontal lines on the graph represent different compute bounds:
\vspace*{-1mm}
\begin{itemize}
    \item SP Vector Add: single precision vector add peak bound (24.06 GFLOPS);
    \item DP Vector Add: double precision vector add peak bound (12.58 GFLOPS);
    \item Scalar Add: scalar add peak bound (3.17 GFLOPS).
\end{itemize}
\vspace*{-1mm}
The data points in the figure represent the computational kernels: $add2$, $add2s1$, $glsc3$, and $ax\_e$ are limited by the L2 memory bound, while the $mxmf2$ (the implementation of matrix multiplication, called by $mxm$) is limited by the DP Vector Add bound.

\begin{figure}[!ht]
    \vspace*{-6mm}
    \centering
    \includegraphics[width=0.95\textwidth]{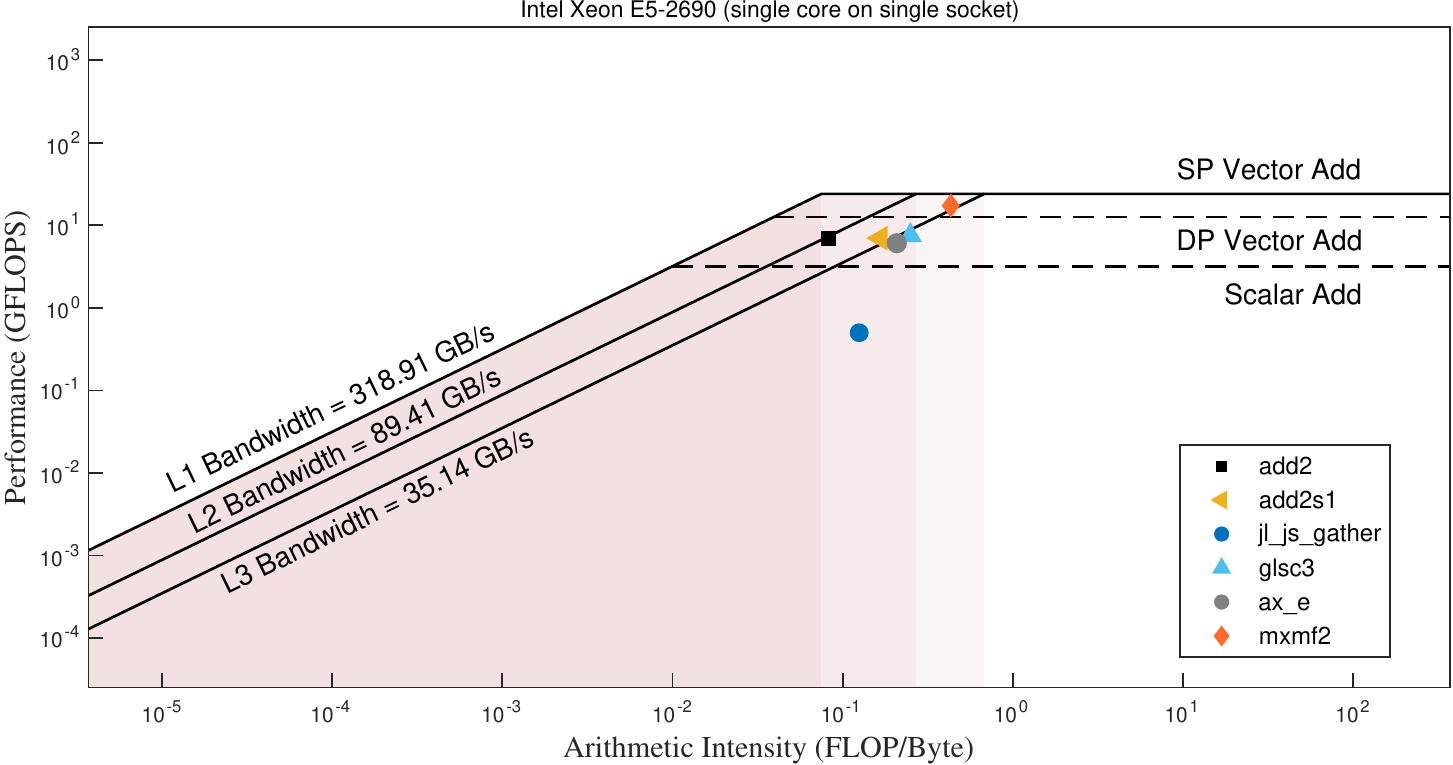}
    \caption{Roofline model for mixed-precision Nekbone.}
    \label{fig:roofline-single}
    \vspace*{-10mm}    
\end{figure}

The implementation of the mixed-precision Nekbone, the same test case, has led to a notable improvement, see~\Cref{fig:roofline-single}: $mxmf2$ surpassed the limits of DP Vector Add bound, with the remaining kernels break the L2 memory bound. \Cref{tab:roofline-dp-sp} presents a comparison between the original and mixed-precision versions.
\begin{table}[!ht]
\vspace*{-6mm}
\caption{Roofline bounds, double and mixed-precision.}
\label{tab:roofline-dp-sp}
\centering
\begin{tabular}{p{0.15\textwidth}||p{0.30\textwidth}p{0.27\textwidth}}
\hline \noalign{\hrule height 0.3pt}
& Double Precision & Mixed Precision \\
\hline 
add2 & L2 memory bound & SP Vector Add bound \\
add2s1 & L2 memory bound & L3 memory bound \\
glsc3 & L2 memory bound & L3 memory bound \\
ax\_e & L2 memory bound & L3 memory bound \\
jl\_gs\_gather & L1 memory bound & Scalar Add bound \\
mxmf2 & DP Vector Add bound & SP Vector Add bound \\
\hline \noalign{\hrule height 0.3pt}
\end{tabular}
\vspace*{-6mm}
\end{table}

\section{A strategy to enable mixed-precision in Nekbone}
\label{sec:mixed}
Implementing mixed-precision in Nekbone cannot be simplified to automatically find and replace data formats or just relying on compiler options. For instance, we tried to impose single precision instead of double precision in the entire code during the compilation of Nekbone, leading to no success. After a thorough analysis of the codes, we have identified the following pitfalls:\\[-4mm]
\begin{enumerate}
    \item Multiple data types are mixed in Nekbone: {\tt REAL}, {\tt REAL*4}, {\tt REAL*8}, {\tt DOUBLE PRECISION}, mainly because of the inconsistent style of type declarations.
    \item Nekbone is designed as a double precision program that requires the use of the compiler options {\tt -fdefault-real-8} ({\tt gfortran, flang}) or {\tt -r8} ({\tt ifort}) for its compilation, these ensure that all variables declared as {\tt REAL}, {\tt REAL*8} and {\tt DOUBLE PRECISION} are treated as double precision.
    \item This inconsistency makes the option {\tt -freal-8-real-4} fail with {\tt gfortran}, used to demote {\tt REAL} types with a word length of 8 to 4. Also, this option is not compatible with {\tt ifort} and {\tt flang} as there is no such compiler flag.
    \item As {\tt implicit none} is a characteristic of modern Fortran, implicit types are permitted in F77, this may result in unexpected behaviours in a mixed-precision environment.
\end{enumerate}
\vspace*{-1mm}

For the second of these pitfalls, {\tt REAL} is still treated as single precision in a double precision environment (with {\tt -fdefault-real-8}), this enables minimal control over the single precision components. Based on results of code inspection with Verificarlo and the consideration of these pitfalls, our strategy for implementing mixed-precision in Nekbone is based on using  single precision in the CG loop only and reducing type conversions to minimise additional overhead.

Practically, we implement the mixed-precision Nekbone in the following steps:\\[-4mm]
\begin{enumerate}
    \item Identify all the modules involved in the (preconditioned) CG computational kernels based on the profiling results.
    \item Explicitly declare all implicit variables as {\tt REAL*4} type, rewrite the parameters and computation parts in single precision (including: \verb|math.f|, \verb|mxm_*.f|, \verb|proc_*.f| and \verb|cg.f|).
    \item Maintain the initialisation part in double precision, only perform type conversion on the variables used within the CG loop.
    \item For the preconditioner part, maintain the initialisation also in double, while using single precision for the computation (including: \verb|hsmg_*.f|).
    \item The C interface in Gather-Scatter library inherently supports single precision, by modifying the Fortran interface to allow single precision communications (including: \verb|mpi_*.f| and \verb|comm_mpi.f|).
\end{enumerate}

\section{Results and performance}
\label{sec:results}
We evaluate the performance of the mixed-precision Nekbone in terms of three dimensions: accuracy, time-to-solution, and energy-to-solution. We set the CG loop stopping criteria to $1.0E-10$. The codes is compiled with {\tt flang} v.14.0.1 and the {\tt -O2} optimisation level on the Kebnekaise Skylake nodes equipped with two Intel Xeon Gold 6132 CPUs with 14 cores @2.6~GHz and Cray Fortran v16.0.1 with {\tt -O2} on the LUMI-C partition equipped with two AMD EPYC 7763 CPUs with 64 cores @2.45~GHz.
\subsection{Accuracy}
Absolute error ($AE$) and mean absolute error ($MAE$) are introduced here for evaluating the accuracy:
\vspace*{-3mm}
$$AE=\left| \Delta residual\right| =\left| r_{m} -r_{d} \right| \quad\quad MAE=\frac{1}{n}\sum _{i=1}^{n} \left|\Delta residual_{i} \right|$$\\[-4.5mm]
where $r_{m}$ is the computed residual of mixed-precision and $r_{d}$ is the computed residual of double precision.

\begin{figure}[!ht]  
    \vspace*{-8mm}
    \begin{subfigure}{0.49\textwidth}
        \includegraphics[width=1.05\linewidth]{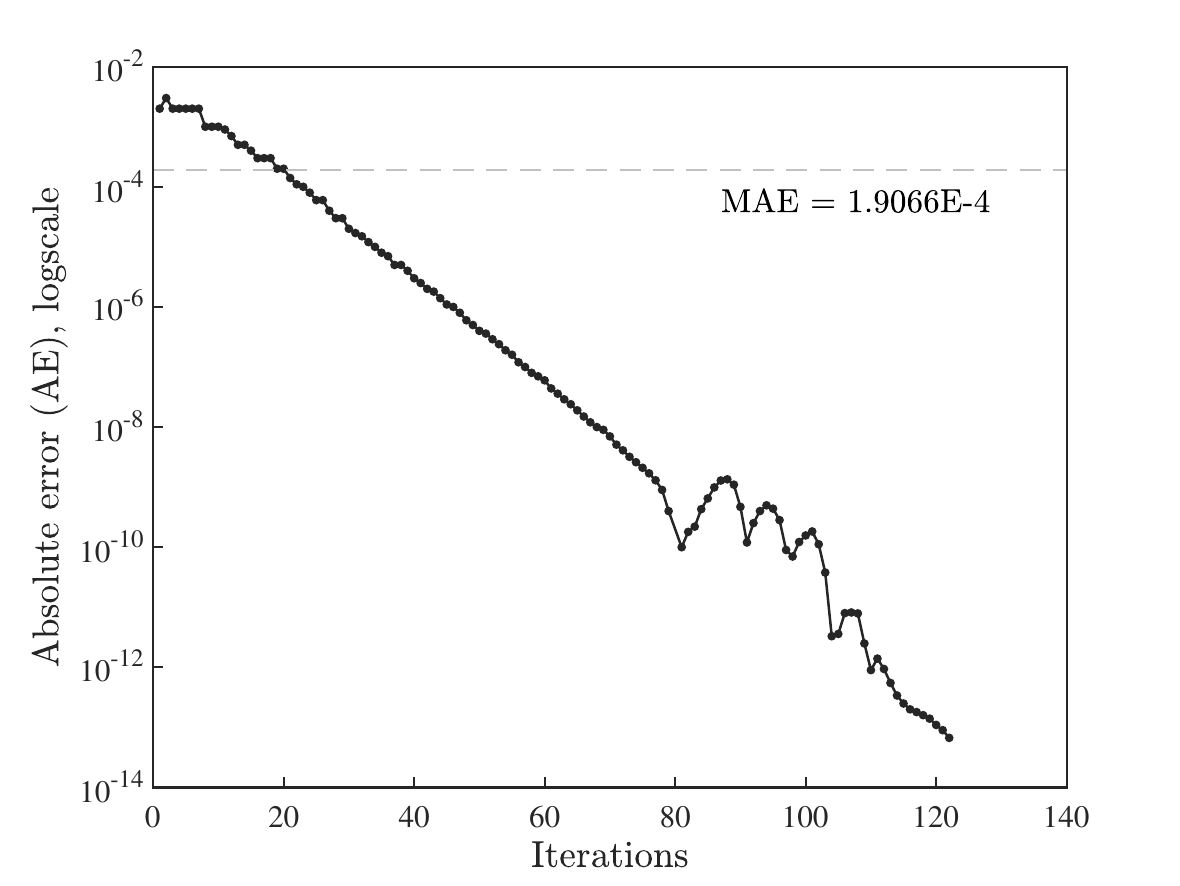}
        \caption{Without preconditioner} \label{fig:ae-noprecond}
    \end{subfigure}
    \hspace*{\fill}
    \begin{subfigure}{0.49\textwidth}
        \includegraphics[width=1.05\linewidth]{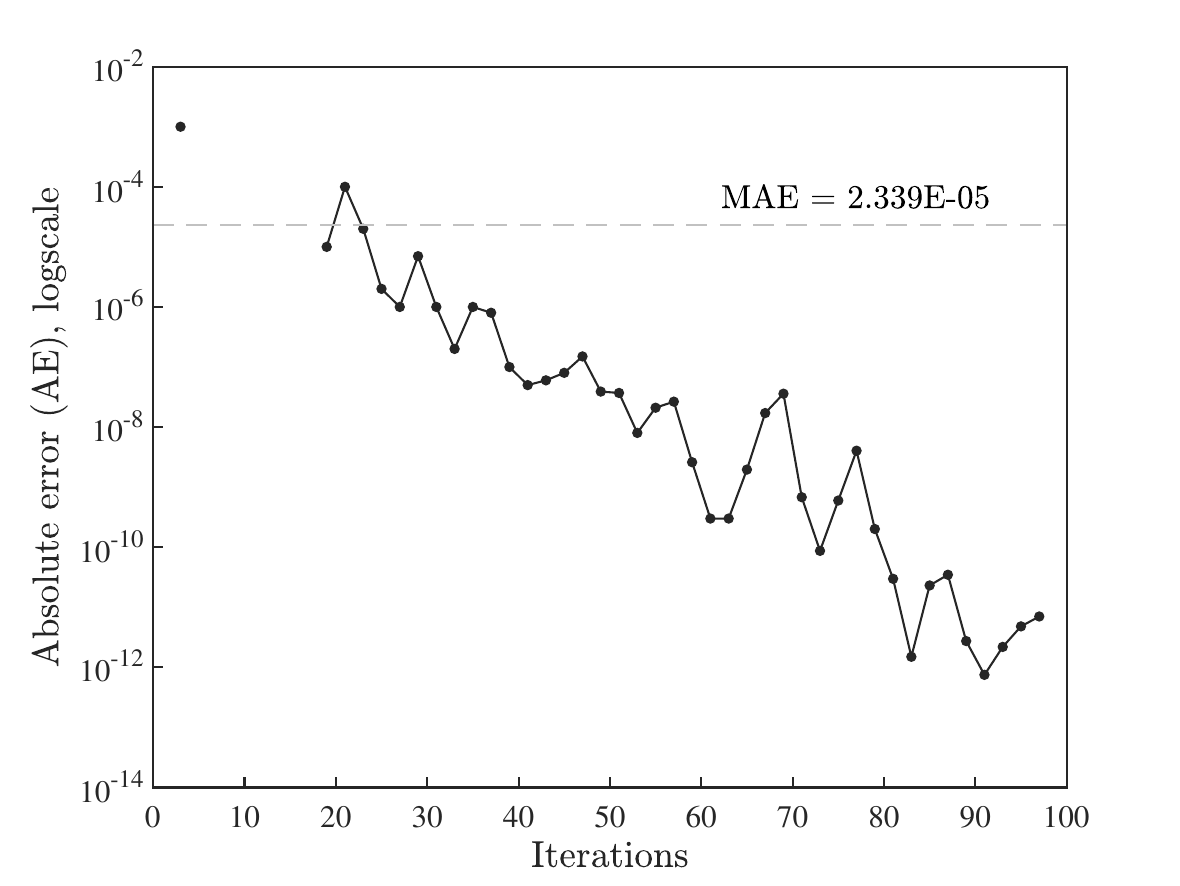}
        \caption{With preconditioner} \label{fig:ae-precond}
    \end{subfigure}
    \vspace*{-6mm}
    \caption{Absolute error (AE) of the mixed-precision against double version.}
    \label{fig:accuracy-ae}
    \vspace*{-8mm}
\end{figure}

\Cref{fig:accuracy-ae} illustrates the history of the absolute error with respect to the number of iterations. For the test case without preconditioner, the error is reducing over the course of iterations and reaches the tolerance at ($iteration=122$) with the value of $6.70E-14$, with the $MAE$ of $1.91E-4$. When the preconditioner is enabled, the absolute error is decreasing with some fluctuations, the error at convergence ($iteration=97$) is $7.03E-12$ with the $MAE$ of $2.34E-5$; for several initial iterations the error is zero and is not visible on~\Cref{fig:ae-precond}.

\begin{figure}[!ht]
    \vspace*{-3mm}
    \begin{subfigure}{0.49\textwidth}
        \includegraphics[width=1.05\linewidth]{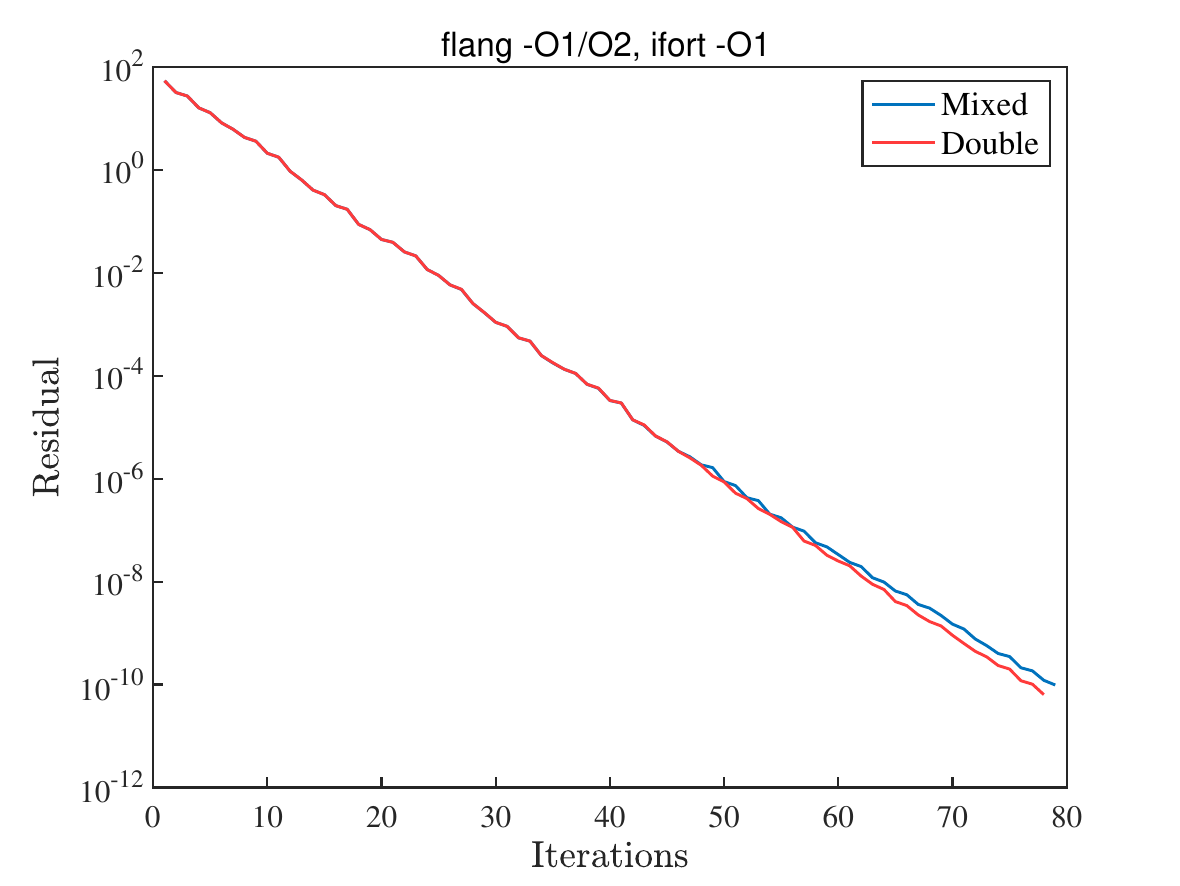}
        \caption{{\tt flang -O1/ -O2}, {\tt ifort -O1}} \label{fig:stagnation-O1}
    \end{subfigure}
    \hspace*{\fill}
    \begin{subfigure}{0.49\textwidth}
        \includegraphics[width=1.05\linewidth]{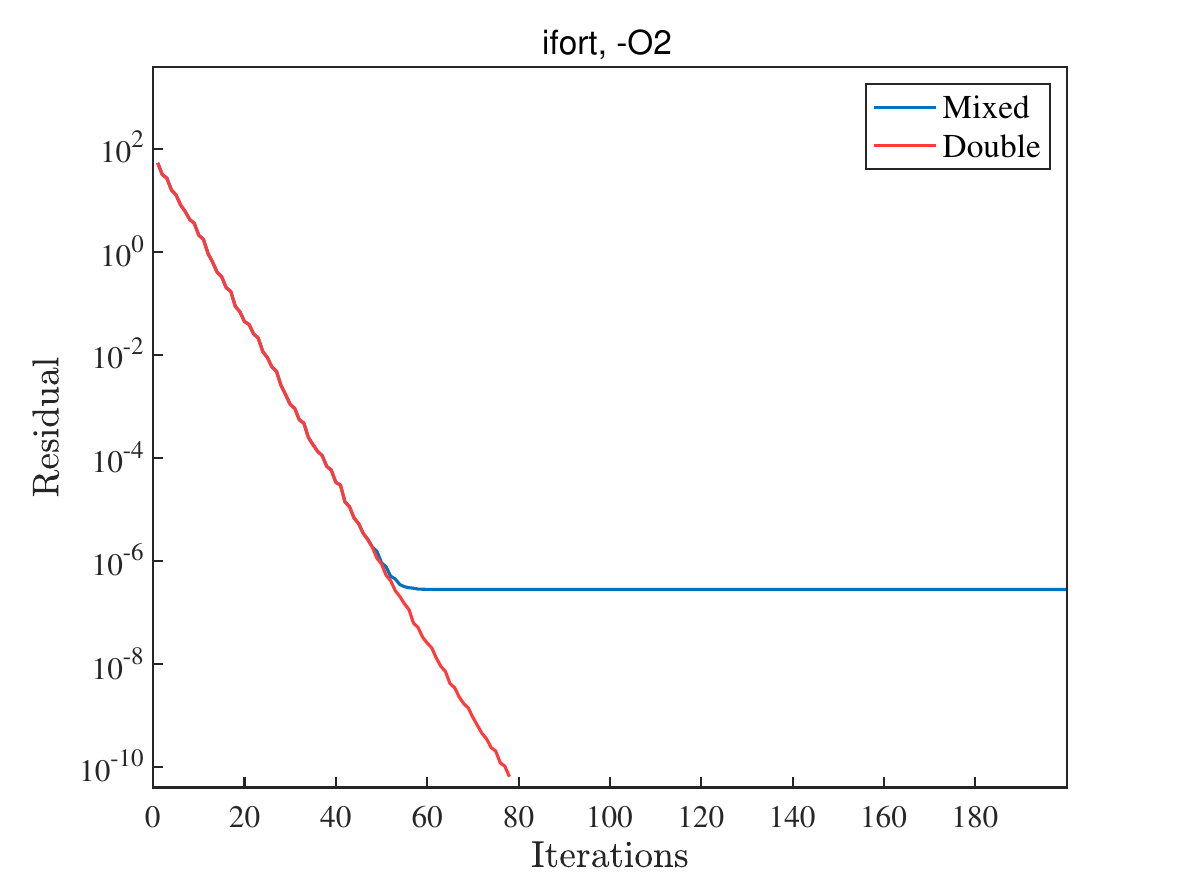}
        \caption{{\tt ifort -O2}} \label{fig:ifort-O2}
    \end{subfigure}
    \vspace*{-6mm}
    \caption{Residual history with {\tt flang \& ifort} and two optimisation levels.}
    \label{fig:stagnation-compiler}
    \vspace*{-7mm}
\end{figure}

When measuring accuracy of the mixed-precision Nekbone with the preconditioner enabled, we observe stagnation with {\tt ifort} and the optimisation level {\tt -O2}, see~\Cref{fig:ifort-O2}. This behavior was predicted during the initial inspection phase by the Verificarlo MCA backend, see~\Cref{fig:mca-precond}.  With {\tt flang} and the optimisation levels of {\tt -O1} and {\tt -O2} and {\tt ifort -O1} there is no stagnation, as depicted in~\Cref{fig:stagnation-O1}. However, the stagnation start to appear even with {\tt flang} after using four MPI ranks. We leave this for further investigation and foresee to introduce precision increase/ adaptivity once required. For a note, there is no stagnation for the case without preconditioner. 

\subsection{Time-to-solution}
We conduct tests on both serial (not compiled with MPI) and parallel program. For the serial tests, we set the number of elements per process to $16$, $50$, and $128$, respectively. We measure two types of program runtime: 1/ the elapsed time of the whole program, counted with the {\tt time} command in GNU/ Linux; 2/ the solve time (the CG loop time), counted by Nekbone internal interfaces that is also included as a part of the output. We report median of five runs.

\begin{figure}[ht]
    \vspace*{-5mm}
    \centering
    \includegraphics[scale=0.45]{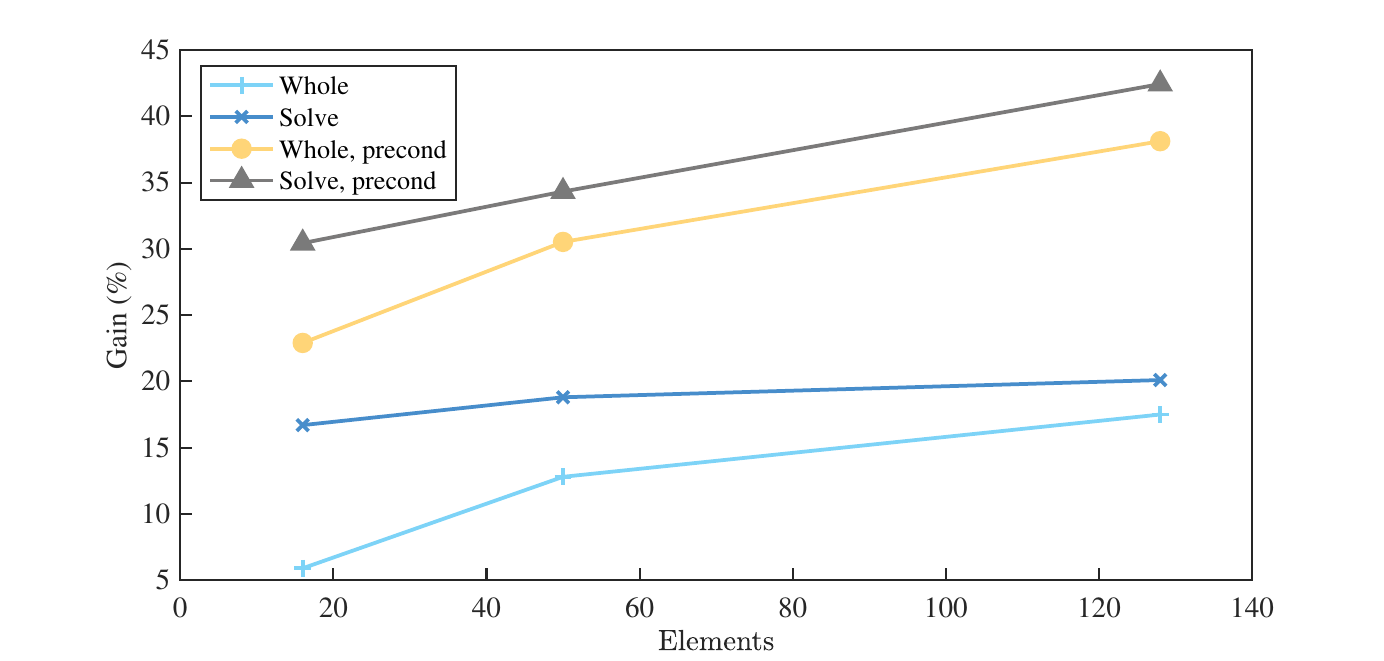}
    \caption{Mixed-precision gain in execution time for different elements.}
    \label{fig:time-serial}
    \vspace*{-6mm}
\end{figure}

\begin{table}[!ht]
    \vspace*{-9.5mm}
    \caption{Elapsed time (seconds) of Nekbone, without preconditioner.}
    \vspace*{-2mm}    
    \begin{subtable}{.49\linewidth}
      \centering
        \caption{Whole program}
        \begin{tabular}{p{0.23\textwidth}||p{0.16\textwidth}p{0.16\textwidth}p{0.16\textwidth}}
        \hline \noalign{\hrule height 0.3pt}
         Elements & 16 & 50 & 128 \\
        \hline 
         Mixed & 0.096 & 0.273 & 0.669 \\
         Double & 0.102 & 0.313 & 0.811 \\
         Gain & 5.9\% & 12.8\% & 17.5\% \\
        \hline \noalign{\hrule height 0.3pt}
        \end{tabular}
    \label{tab:time-serial-noprecond-whole}
    \end{subtable}%
    \begin{subtable}{.49\linewidth}
      \centering
        \caption{Solve time}
        \begin{tabular}{p{0.23\textwidth}||p{0.16\textwidth}p{0.16\textwidth}p{0.16\textwidth}}
        \hline \noalign{\hrule height 0.3pt}
         Elements & 16 & 50 & 128 \\
        \hline 
         Mixed & 0.035 & 0.117 & 0.303 \\
         Double & 0.042 & 0.144 & 0.379 \\
         Gain & 16.7\% & 18.8\% & 20.1\% \\
        \hline \noalign{\hrule height 0.3pt}
        \end{tabular}
    \label{tab:time-serial-noprecond-solve}
    \end{subtable}
\label{tab:time-serial-noprecond}
    \vspace*{-7mm}
\end{table}
The timing results of the serial program are shown in~\Cref{tab:time-serial-noprecond} for the test case without preconditioner and in~\Cref{tab:time-serial-precond} for the case with preconditioner. The gain is calculated by $Gain=\dfrac{T_{double} -T_{mixed}}{T_{double}} \times 100\%$. For the case without preconditioner, the gain increases with the number of elements. The mixed-precision version of Nekbone demonstrates significant advantage especially when focusing on the solve time. When the preconditioner is enabled, the mixed-precision shows a greater advantage in both the whole program and the solve time: the gain reaches $42.43\%$ for $128$ elements. \Cref{fig:time-serial} illustrates benefits of the mixed-precision Nekbone in terms of the reduced time-to-solution for various number of elements for both cases. The gain is growing with the number of elements. Notably, the trend is stronger in the solve part and for the case with preconditioner that has roughly 2x gain compared to the case with the CG only.

\begin{table}[!ht]
    \vspace*{-9.5mm}
    \caption{Elapsed time of Nekbone, with preconditioner.}
    \vspace*{-3mm}
    \begin{subtable}{.49\linewidth}
      \centering
        \caption{Whole program}
        \begin{tabular}{p{0.225\textwidth}||p{0.16\textwidth}p{0.16\textwidth}p{0.16\textwidth}}
        \hline \noalign{\hrule height 0.3pt}
         Elements & 16 & 50 & 128 \\
        \hline 
         Mixed & 0.165 & 0.560 & 1.630 \\
         Double & 0.214 & 0.806 & 2.634 \\
         Gain & 22.90\% & 30.52\% & 38.12\% \\
        \hline \noalign{\hrule height 0.3pt}
        \end{tabular}
    \label{tab:time-serial-precond-whole}
    \end{subtable}%
    \begin{subtable}{.49\linewidth}
      \centering
        \caption{Solve time}
        \begin{tabular}{p{0.225\textwidth}||p{0.16\textwidth}p{0.16\textwidth}p{0.16\textwidth}}
        \hline \noalign{\hrule height 0.3pt}
         Elements & 16 & 50 & 128 \\
        \hline 
         Mixed & 0.064 & 0.243 & 0.711 \\
         Double & 0.092 & 0.370 & 1.235 \\
         Gain & 30.43\% & 34.32\% & 42.43\% \\
        \hline \noalign{\hrule height 0.3pt}
        \end{tabular}
    \label{tab:time-serial-precond-solve}
    \end{subtable}
\label{tab:time-serial-precond}
    \vspace*{-4mm}
\end{table}

For parallel tests without preconditioner, we performed weak scaling tests on one node on LUMI-C with the fixed number of elements $128$ and vary the number of MPI ranks. \Cref{tab:time-mpi-noprecond} reports time-to-solution and highlights the mixed-precision gain. The gain is increasing with the number of processes resulting in 40.68\,\% for the whole program and 61.77\,\% for the solve on 128 MPI ranks.
\begin{table}[!ht]
\vspace*{-9.5mm}
\centering
\caption{Elapsed time (secs) with $128$ elements per MPI rank on LUMI-C.}
\vspace*{-2mm}
\begin{subtable}{\linewidth}
    \centering
    \caption{Whole program}
    \begin{tabular}{p{0.137\textwidth}||p{0.08\textwidth}p{0.08\textwidth}p{0.08\textwidth}p{0.09\textwidth}p{0.09\textwidth}p{0.09\textwidth}p{0.09\textwidth}}
    \hline \noalign{\hrule height 0.3pt}
    MPI ranks & 1 & 4 & 8 & 16 & 32 & 64 & 128 \\
    \hline 
    Mixed & 0.741 & 0.905 & 0.995 & 1.642 & 1.694 & 2.096 & 2.641 \\
Double & 0.775 & 1.026 & 1.857 & 2.920 & 3.151 & 3.562 & 4.452 \\
Gain & 4.39\% & 11.79\% & 46.42\% & 43.77\% & 46.24\% & 41.16\% & 40.68\% \\
    \hline \noalign{\hrule height 0.3pt}
    \end{tabular}
    \label{tab:time-mpi-noprecond-whole}
\end{subtable}
\vspace*{-4mm}
\begin{subtable}{\linewidth}
    \centering
    \caption{Solve time}
    \begin{tabular}{p{0.137\textwidth}||p{0.08\textwidth}p{0.08\textwidth}p{0.08\textwidth}p{0.09\textwidth}p{0.09\textwidth}p{0.09\textwidth}p{0.09\textwidth}}
    \hline \noalign{\hrule height 0.3pt}
    MPI ranks & 1 & 4 & 8 & 16 & 32 & 64 & 128 \\
    \hline 
    Mixed & 0.165 & 0.178 & 0.190 & 0.256 & 0.430 & 0.445 & 0.476 \\
    Double & 0.182 & 0.239 & 0.596 & 1.115 & 1.161 & 1.207 & 1.245 \\
    Gain & 9.34\% & 25.52\% & 68.12\% & 77.04\% & 62.96\% & 63.13\% & 61.77\% \\
    \hline \noalign{\hrule height 0.3pt}
    \end{tabular}
    \label{tab:time-mpi-noprecond-solve}
\end{subtable}
\label{tab:time-mpi-noprecond}
\vspace*{-2mm}
\end{table}

\subsection{Energy-to-solution}
Measuring energy consumption is not so easy as time especially since it is not widely supported. Due to this, we use LUMI-C that supports energy measurements with the exclusive use of a single node. We made the performance results in~\Cref{tab:time-mpi-noprecond} coherent with the energy results shown here. In order to measure energy-to-solution of our mixed-precision application, we opted to use the energy accounting plugin provided by Slurm, owing to its simplicity.
We confirm through {\tt slurm.conf} that the underlying mechanism employed by SLURM for energy data collection is {\tt pm\_counters}. This measures power values via a Baseboard Management Controller (BMC) at a frequency of 10 Hz and utilizes this data to calculate the energy consumption for each node. So, it is crucial to have an exclusive use of the node; more details are in~\cite{crayx30powermonitorandmanage}. 
The energy consumption was retrieved after the job has finished with {\tt sacct} using the {\tt --format} option with the desired field {\tt ConsumedEnergy} and specifying the job id. \Cref{tab:energy-mpi} reports the  consumed energy in joules of the entire Nekbone with double and mixed-precision versions. We run each test five times, compute mean as well as the standard deviation (stddev). The {\em reduction in energy-to-solution for the mixed-precision version correlates with the time-to-solution and notably shows better gain}, confirming the efficiency our methodology.

\begin{table}[ht]
    \vspace*{-10mm}
    \centering
    \caption{Energy-to-solution (joules) of Nekbone w various MPI ranks on LUMI-C.}
    \begin{tabular}{p{0.16\textwidth}||p{0.11\textwidth}p{0.11\textwidth}p{0.11\textwidth}p{0.11\textwidth}p{0.11\textwidth}p{0.08\textwidth}}
    \hline \noalign{\hrule height 0.3pt}
    MPI ranks & 32 & stddev & 64 & stddev & 128 & stddev \\
    \hline 
    
    Mixed  & 451.6   &4.1\%& 653.6 &2.4\% & 1089.4   &1.0\%\\
    Double & 990.6   &3.9\%& 1424.8 &4.5\%& 2061.2   &3.2\%\\
    Gain   & 54.41\% & &54.12\% & & 47.14\% & \\
    \hline \noalign{\hrule height 0.3pt}
    \end{tabular}
    \label{tab:energy-mpi}
    \vspace*{-10mm}
\end{table}

\section{Conclusion and Future Work}
\label{sec:conclusion}
We have presented a systematic methodology for enabling mixed-precision, which consists of three main parts: inspection with the help of tools, implementation, and validation. Using the Nekbone as a case study, we introduce mixed-precision in the solve part for the two main test cases, namely with and without preconditioner. The mixed-precision version reduces time-to-solution by 40.7\,\% and energy-to-solution by 47.1\,\% on 128 MPI ranks. 

In the near future, we will explore adaptivity in the CG solver to increase or lower precision when there is such a need or an opportunity. For instance, for the case with preconditioner where, we observe stagnation in convergence after $7.94\times 10^{-6}$ that is aligned with Verificarlo prediction. However, we also would like to argue a need of high accuracy in solvers used in scientific codes which should not exceed the discretisation error. 
In addition, this work will be extended to large applications like Neko and SOD2D from the CEEC project.

\subsubsection*{Acknowledgment} We would like to thank G\"ulçin Gedik for her help in measuring energy. This research was partially supported by a Center of Excellence in Exascale CFD (CEEC) grant No 101093393 from \worldflag[width=3.2mm]{EU} via EuroHPC JU and \worldflag[width=3mm]{SE}\worldflag[width=3mm]{DE}\worldflag[width=3mm]{ES}\worldflag[width=3mm]{GR}\worldflag[width=3mm]{DK}. We acknowledge NAISS \worldflag[width=3mm]{SE}, partially funded by VR no. 2022-06725, for access to LUMI, EuroHPC JU, and hosted by CSC \worldflag[width=3mm]{FI}.
\vspace*{-2mm}

\bibliographystyle{splncs04}
\bibliography{references}

\end{document}